**DETECTION OF NORTH ATLANTIC RIGHT WHALE UPCALLS USING**

**LOCAL BINARY PATTERNS IN A TWO-STAGE STRATEGY**

Mahdi Esfahanian, mesfahan@fau.edu*

Hanqi Zhuang, zhuang@fau.edu

Nurgun Erdol, erdol@fau.edu

Edmund Gerstein, egerste1@fau.edu

Department of Computer and Electrical Engineering and Computer Science, Florida Atlantic

University, Boca Raton, FL 33431, USA

* Corresponding author. Tel: +1 561 929 6392

    E-mail address: mesfahan@fau.edu

**Abstract**

In this paper, we investigate the effectiveness of two-stage classification strategies in detecting north Atlantic right whale upcalls. Time-frequency measurements of data from passive acoustic monitoring devices are evaluated as images. Vocalization spectrograms are preprocessed for noise reduction and tone removal. First stage of the algorithm eliminates non-upcalls by an energy detection algorithm. In the second stage, two sets of features are extracted from the remaining signals using contour-based and texture based methods. The former is based on extraction of time-frequency features from upcall contours, and the latter employs a Local Binary Pattern operator to extract distinguishing texture features of the upcalls. Subsequently evaluation phase is carried out by using several classifiers to assess the effectiveness of both the contour-based and texture-based features for upcall detection. Experimental results with the data set provided by the Cornell University Bioacoustics Research Program reveal that classifiers show accuracy improvements of 3% to 4% when using LBP features over time-frequency features. Classifiers such as the Linear Discriminant Analysis, Support Vector Machine, and TreeBagger achieve high upcall detection rates with LBP features.

## 1. Introduction

The North Atlantic Right Whale (NARW), (*Eubalaena glacialis*), is one of the critically endangered species with an estimated count of fewer than 400 individuals in the North Atlantic Ocean [1, 2]. Right whales have been the "right" whale to hunt leading to their extreme low numbers. North Atlantic right whales can be found in coasts of U.S. and Canada ranging from Bay of Fundy in Canada in the summer to Florida and Georgia coasts in the winter [3, 4]. Since their habitant regions are contaminated by human activities such as shipping traffic and fishing vessels, the anthropogenic mortality from collision with ships and entanglements in fishing gear are considered the main causes of their low population [1]. Therefore, many long-term studies have been conducted with the goal of monitoring NARWs in their habitats, especially where anthropogenic activities have increased. Results of such studies are also useful to marine biology research in conservation and behavioral changes.

North Atlantic right whales produce a variety of vocalizations but our focus is mainly on one of the most commonly heard sounds known as "upcalls" or "contact call" which is characterized by an upsweep frequency from 50 to 350 Hz [5]. These are stereotyped frequency-modulated (FM) calls, about a second in duration. Their function is believed to establish and maintain contact between right whales. Variability of these contact calls is significant due to changes in initial frequency, FM rate, duration and bandwidth. Since upcalls are very common in their vocal repertoire and are highly species-specific, they are used as the primary basis for the acoustical detection of right whales [5, 6]. Other NARW vocalizations described in the literature as tonals, gunshots, hybrid, pulsive, and downcall [7] also exist but their proportion varies by season and habitat.

Generally speaking, the dominant thought has been that the most accurate method for detecting NARW calls in large data sets is to employ human operators to evaluate data spectrograms visually and corroborate the results aurally. This method not only requires a great deal of time and can incur large labor costs, it is also limited by operator judgment which is often subjective and may include false detections. In recent years, Passive Acoustic Monitoring (PAM) [8, 9] has received wide acclaim as one of the most effective techniques for detecting and localizing marine mammals. Acoustic data are also a fundamental component of studying the behavior of many cetaceans.

One of the challenges that arise in using PAM systems is the level of ambient noise which can vary considerably over the course of data collection. This can make the data analysis difficult especially in low SNR environments. Another problem in the design of automatic detectors is the lack of *a* priori knowledge about noise and signals. The probability density function and the power spectral density (PSD) of signals and noise are both unknown. In addition, frequency components of upcalls may change depending on the location, season, time of the day, sex and so on. Another issue in detection of NARW upcalls is the presence of other marine mammals' vocalizations in the recordings. The sound most similar to the right whale upcall which causes the greatest problem is that of the humpback whales (*Megaptera novaeangliae*). The situation gets even more complicated since the population of humpback whales is more than that of right whales and they vocalize more often and louder.

The detection of NARW upcalls using PAM systems has attracted researchers in the field of bioacoustics since these species are highly endangered and automatic detection systems have to be developed in order to find right whale calls amid other marine mammal vocalizations. Mellinger [10] compared spectrogram correlation and neural network approaches to detection of NARW upcalls. Spectrogram correlation used synthetic kernels designed for right whale upcalls and found the optimal set of parameters over all the available calls. The neural net method trained a set of weights via backpropagation on 9/10 of the test dataset. The neural network performed better, achieving an error rate of less than 6%. Munger *et.al* [11] used the software program *Ishmael* to implement spectrogram cross-correlation with a synthetic kernel [12] for detecting right whale calls [13]. Despite the fact that this detector had high number of false detections and missed individual calls, spectrogram cross-correlation helped human analysts identify segments of data with high probability of containing right whale calls. Another detection procedure reported by Gillespie [14] has two stages. In the first stage parameters are extracted from contours obtained from the smoothed spectrogram using an edge detection method. In the second stage, these parameters are fed into a classifier to find the sounds associated with right whales. This problem was also addressed by Urazghildiiev *et.al* [15] who used a generalized likelihood ratio test (GLRT) detector of polynomial-phase signals with unknown amplitude and polynomial coefficients observed in the presence of locally stationary Gaussian noise. The closed form representation for minimally sufficient statistics was derived and a realizable detection scheme was developed. Urazghildiiev and Clark [16] also designed an automatic detector for a passive acoustic NARW monitoring system where the detector determines the time of signals' probable occurrence but a human operator makes the final decision after inspecting the marked areas of the spectrogram.

In this paper, a new scheme is proposed for NARW upcall detection. In this scheme, after preprocessing, a two-stage operation is performed. In the first stage, obvious non-upcalls are eliminated by an energy detection algorithm. In the second stage features of the remaining signals, which may contain either upcalls or competing non-upcalls, are classified by a binary classifier. To this end, two feature extraction algorithms are studied for their effectiveness for NARW upcall detection. The first method performs elaborate pre-processing in order to isolate a contour associated with an upcall in the spectrogram and then relevant time-frequency parameters are extracted from the contour for classification. The second approach, unlike the aforementioned method, does not rely on contour extraction [17]. It identifies first high intensity regions of the spectrogram. Then a Local Binary Pattern (LBP) operator is applied in the region of interest to capture important texture features. Finally, both types of features are separately fed to classifiers and detection/classification results are evaluated.

The remainder of the paper is organized as follows. Section 2 outlines the proposed detection scheme. Section 3 discusses the feature extraction algorithms, including both the contour-based and texture-based LBP algorithms. Section 4 presents experimental results with a large database of NARW upcalls and other sound files. The paper ends with concluding remarks in Section 5.

## 2. Proposed NARW Upcall Detection Scheme

The two-stage NARW Upcall Detection Scheme used in this work is depicted by the flowchart in Fig. 1. The detection environment can be described in two distinct settings. Examples of recordings with only ambient noise and no marine mammal vocalizations are shown in Fig.2. They are easily determined in the first stage to contain no upcalls. Discrimination of NARW upcalls from competing and very similar vocalizations such as upsweeps produced by Humpback whales is very challenging and requires the more elaborate classifier methods of stage. Fig. 3 shows examples of calls that are indistinguishable from NARW upcalls even by human operators. In the two-stage procedure, those signal segments that do not contain any calls are sorted out first using a simple energy-based detector [10], leaving those more difficult ones for further processing. The elimination of this class of data segments has the clear advantage of reducing the overall computational cost of the process simply by not requiring the more computationally involved classifier stage. Another advantage is that it balances the data going into the classifiers. Balanced data sets increase the effectiveness of such binary classifiers as Support Vector Machine (SVM) and enable their full potential [18].

## 3. Feature Extraction Algorithms

In the second stage, a classification scheme is used for call detection. In the heart of a classifier, resides a feature extraction procedure. The two feature extraction approaches investigated are described below.

### 3.1. Contour-based approach

For contour extraction [19-22], NARW acoustic data sampled at 2 KHz were segmented into 128 ms (256 sample) frames overlapping by 80%. Each frame was transformed into the frequency domain by a 256-point Fast Fourier Transform (FFT) and spectrograms were built with a time resolution of 51 ms. and a frequency resolution of 7.8125 Hz. The spectrogram was normalized according to

$$S_N(t, f_i) = \frac{S(t, f_i) - \mu_i}{\sigma_i} \quad for \quad i = 1, ..., N \tag{1}$$

where $S_N(t,f)$ and $S(t,f)$ represent the normalized and original spectrograms, respectively; $\mu_i$ and $\sigma_i$ are mean and standard deviation, respectively, calculated for each frequency band $f_i$. Such normalization ebbs the long-lasting narrowband noises made by ships, wind and electrical machineries and emphasizes short-duration sounds such as upcalls. For further reduction of background noise and elimination of extreme values, the spectrogram is also equalized by hard-limiting the spectrogram between the upper bound $S_{ceiling}$ and the lower bound $S_{floor}$ by

$$S_H(t, f) = \max\left[ S_{floor}, \min\left( S_{ceiling}, S_N(t, f) \right) \right] . \tag{2}$$

The equalized spectrogram is converted to a binary image with the goal of finding continuous objects. Considering the fact that some upcalls may be very faint inside the background noise in the spectrogram, it is worthwhile to point out the importance of a well-chosen low threshold that has to be set in such a way that no parts of an up-call are missed during the binarization process.

The binary image obtained from the equalized spectrogram of Fig. 4 is given in the top image of Fig. 5.

The same binary image shows many spurious contours remaining as a result of the tradeoff for choosing a low threshold. Ranging from tiny to large, these contours are irrelevant clutter and must be separated from the object of interest corresponding to the upcall contour. Therefore, an 8-connected neighborhood and Moore-Neighbor tracing technique using Jacob's stopping criteria [23] are applied on the spectrogram to locate individual continuous objects and trace their exterior boundaries as shown in the lower image of Fig. 5. Then a set of properties such as perimeter (pixels), area (pixels), height (Hz) as a measure of frequency range, width (sec) as a measure of the time duration, is extracted from each object in the image for further processing. The parameters are used to make an initial decision to discard an object or keep it for further analysis. The thresholds for each object are chosen to minimize the number of missed objects associated with an upcall. Now one of the three cases can happen after the preprocessing phase:

1. No objects are detected, indicating that the spectrogram contains no upcalls and therefore is labeled as no-upcall as Fig. 6.

2. A single object is detected in the spectrogram. In this situation, that object is considered as the potential upcall and will be passed to the second detection phase as depicted in Fig. 7.

3. In the last scenario, two or more objects are detected. This case is usually seen where the intensity of spectrogram pixels at some point along an upcall contour is low and falls below the threshold, which leads to a break along the contour. To tackle this issue, a set of criteria such as minimum and maximum length of each object in time and frequency as well as frequency and time spacing between two objects are utilized. If these criteria are met, the algorithm will merge them together and consider it as a potential upcall in the next detection phase. On the other hand if the criteria are not met, the objects are considered as separate ones.

The first stage of detection categorizes the audio signal segments in which no object is found into the "non-upcall" class. Any signal segment which does not belong to the 'non-upcall' class is fed to the second stage to determine if there is an upcall in the segment. At this point, a feature vector has to be computed for all objects considered as potential upcalls. For this purpose, a set of features named "TFP-2 features" are extracted from detected objects. These features are defined as follows: minimum frequency (Hz), maximum frequency (Hz), frequency band (Hz), perimeter (pixels), area (pixels), orientation (degree) and time duration (sec).

### 3.2. Texture based approach

Most of the upcalls in our data set have been observed to occur within the frequency range of 80 Hz and 320Hz, respectively. Hence, the first step of preprocessing in this approach involves a band-pass filter that limits the frequency range of NARW calls. The next step which is similar to that of the first method is to run normalization and equalization algorithms on the spectrogram in order to enhance the upcalls and remove clutter. After preprocessing, the LBP algorithm as described in appendix, is applied to extract features in the spectrogram.

LBP is a feature extraction method which is capable of describing texture patterns in the image [24,25]. To detect an upcall in the image, the LBP operator scans the entire spectrogram using a circular 8-point neighborhood of radius 1 along with "uniform pattern" concept yielding to the LBP image as depicted in the top image of Fig. 8. It is evident from the image that the upcall contour is preserved after the LBP operation. Feature vectors are subsequently derived from the LBP histogram, as depicted in the bottom image of Fig. 8, following the method described in [24,25].

## 4. Detection Results

In this study, all recordings were made using archival bottom-mounted acoustic recorders designed by the Cornell University Bioacoustics Research Program. These units are referred to as "Pop-ups" because they are released from the bottom by an acoustically triggered release mechanism that allows them to float to the surface for recovery. Each unit consisted of an HTI-94-SSQ hydrophone with a sensitivity of $-168$ dB re 1 V/$\mu$Pa, an amplifier with a gain of 23.5 dB, and an A/D converter with a sensitivity of $10^3$ bit/V. An additional normalization of the A/D converter output by a factor of 1/2048 was implemented when storing the digitized data on a hard drive. The final transformation coefficient used for calibration of the digitized data was $C\tilde{} = -168.5 + 23.5 + 20\log(10^3) - 20\log(2048) = -151.2$ dB re 1 $\mu$Pa. The systems had a flat ($\pm 1.0$ dB) frequency response between 10 and 585 Hz, which included the bandwidth of right whale upcalls (50–350 Hz) used for the measurements reported here [26]. Pop-up units were deployed in one of right whale habitant areas named Cape Cod Bay (CCB) in Massachusetts, a late winter and early spring feeding ground as shown in Fig. 9.

The effectiveness of the proposed approach described in the previous sections is evaluated for NARW upcall detection with various popular classifiers. In the training phase, 4000 right whale audio segments are utilized, in which there are 1265 NARW upcalls and 2735 non-upcalls. In addition, the classification methods for upcall detection are tested on 3000 NARW audio segments in which there are 699 upcalls and 2301 non-upcalls. In this section, three different detection rates are used to analyze detection results:

$$\text{Overal detection rate} = \frac{\text{number of correctly classified calls}}{\text{total number of calls}} \qquad (3)$$

$$\text{Upcall detection rate} = \frac{\text{number of correctly classified upcalls}}{\text{total number of upcalls}} \qquad (4)$$

$$\text{Non-upcall detection rate} = \frac{\text{number of correctly classified non-upcalls}}{\text{total number of non-upcalls}} \qquad (5)$$

Table 1 shows the detection results obtained from various classifiers using TFP-2 features. It is observed that the highest rate of correct detection is achieved by the Linear Discriminant Analysis (LDA) with 80% accuracy(560 of 699 upcalls) followed by TreeBagger at 76% (533 of 699 upcalls) and Linear SVM at 71 % (495 of 699 upcalls). Although Decision Tree and KNN were the wors performers in the detection of NARW upcalls, the best non-upcall detection rate (low false alarm) is also achieved by the Decision Tree at 98.5% accuracy (2267 of 2301 non-upcalls) followed by Linear SVM at 97% (2234 of 2301 non-upcalls). It is worthwhile to mention the high accuracy of linear SVM compared with Gaussian and polynomial kernel SVM methods where the latter struggles specifically in upcall detection with only 42% accuracy. The last column also reveals that LDA, Treebagger, and Linear SVM are the high performers with overall detection rates of 91.7%, 91.27%, and 90.97%, respectively. To summarize the entire classification results, the Receiver Operating Characteristics (ROC) curves which are plots of true positive rate (correctly classified upcall) against false positive rate (non-upcall classified as upcall) are shown for all scenarios in Fig. 10. The closer the ROC curve follows the vertical axis and then the top border of Fig. 10, the more accurate is the classifier. As evidenced in Fig. 10, LDA, Treebagger, and Linear SVM curves achieve high true positive and low false positives hence the area under these ROC curves is greater than others.

Detection results are given in Table 2 using LBP features. In terms of best upcall detection rate, Linear SVM with 90.41% (632 of 699 upcalls) outperforms the other classifiers followed by TreeBagger with 89.98% accuracy (629 of 699 upcalls). On the other hand, LDA is cable of obtaining 98% accuracy in non-upcall detection corresponding to 2251 upcalls. It is also interesting that although KNN is a very simple classifier it can achieve a high non-upcall detection rate (95%) while keeping the upcall detection rate acceptably high (79%). While RBF SVM achieved accurate non-upcall detection rate of 94%, polynomial SVM shows very low upcall detection 59%. The best detection performance is achieved by Linear SVM, TreeBagger, and LDA with accuracies around 92-93%. For the overall comparison of classifiers tested with LBP features, their ROC curves are plotted in Fig. 11 confirming the above claim since these three classifiers have the largest area under the curve. Comparing ROC curves in Fig. 10 and Fig. 11 reveals that classifiers with LBP features have gained about 3% to 4% accuracy improvement over TFP-2 features with an identical classifier. The AUC values computed for all algorithms depicted in Tables 3 and 4 confirm the aforementioned remarks.

## 5. Conclusions

A two-stage scheme for detection of NARWs is proposed in the paper. In the first stage, most non-upcalls are picked up by an energy-based detector. In the second stage, features of NARW upcalls are extracted using both contour and texture based algorithms. Various classifiers such as LDA, SVM, and TreeBagger are paired with these feature extractors and their detection results are analyzed. The detector that acquired the highest accuracy (around 91%) with TFP-2 features is the TreeBagger since this approach creates an ensemble of decision trees where every tree is grown on an independently drawn bootstrap replica of input data. On the other hand, LDA is observed to detect the highest number of upcalls and linear SVM demonstrates the least false negative rate. With only LBP features, LDA, linear SVM, and Tree Bagger are amongst the high-ranked detectors with accuracies close to 93%. It seems that LBP features exhibit linear characteristics that discriminant analysis and SVM with linear kernels can well distinguish upcalls in the dataset. The largest percent of upcall detection (true positive) belong to linear SVM and on the other hand, LDA produces the least number of non-upcall misclassifications. An important observation is that switching from TFP-2 features to LBP features produces considerably high detection rate with all classifiers tested which once again indicate highly-informative property of the LBP features.

## Acknowledgments

The authors acknowledge the financial support of a Seed Grant from Florida Atlantic University and the acquisition of the acoustic data made available by Cornell University.

## Appendix: Local Binary Patterns

The LBP operator [22] serves as a descriptor of local spatial patterns. An LBP label is a binary number created for each pixel on a small neighborhood where each bit is assigned based on its difference from the center of the neighborhood at a given radius. Fig. A.1-a shows a simple version that labels each pixel by a number derived from its 3×3 neighborhood. Neighboring pixels are assigned a binary value of 1 if larger than the center pixel and 0 otherwise. An 8-bit binary number is formed by concatenating the bits and its decimal equivalent becomes the label for the center pixel indicated by a circle. The LBP coder can be extended to allow neighborhoods with different sizes. A circularly symmetric neighborhood can be constructed by defining a two parameter pair $(P, R)$: the number of equally spaced points $P$ controls the quantization of angular space and the radius $R$ determines the spatial resolution of LBP operator. Fig. A.1-b illustrates how bits may be clustered to indicate local primitives such as spots, lines, edges, and corners. Locations that do not have integer coordinates are evaluated by a bilinear interpolation of the neighboring pixels. The feature vectors are obtained from the histograms of the LBP-operated image sections, which are explained next. Let $f_l(x, y)$ be the LBP label of pixel $(x, y)$. The histogram of region $j$ is computed by:

$$H_{i,j} = \sum_{x,y} I\{f_l(x, y) = i\} I\{(x, y) \in R_j\} \tag{A.1}$$

where $i = 0, 1, ..., n-1, j = 0, 1, ..., m-1$ and

$$I\{A\} = \begin{cases} 1 & A \text{ is true}, \\ 0 & A \text{ is false}. \end{cases} \tag{A.2}$$

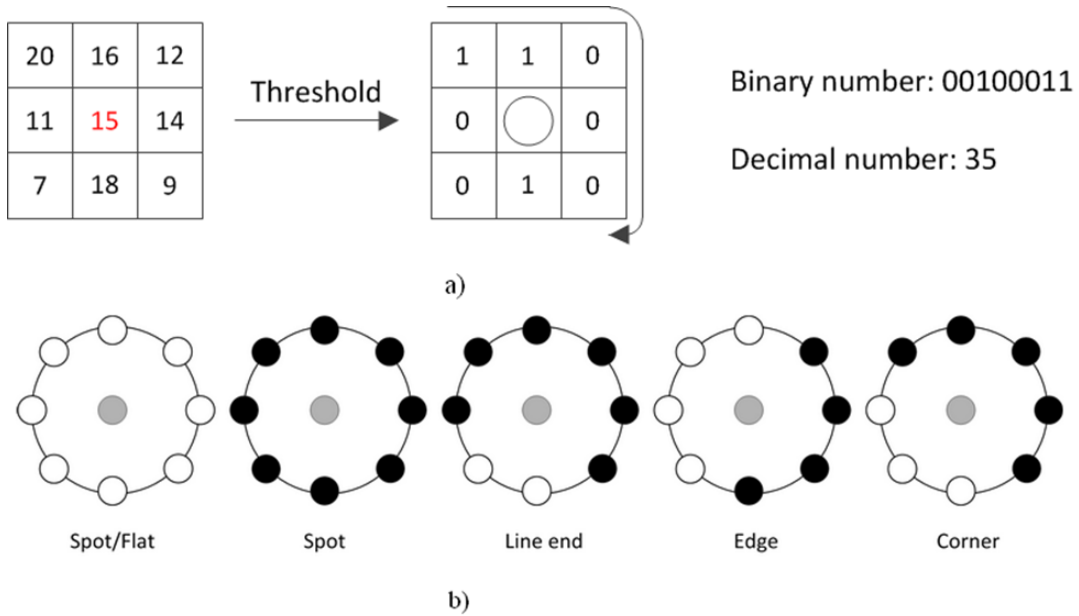

a)

b)

**Fig. A.1.** a) The LBP operator. b) Examples of texture primitives which can be detected by LBP, where white circles represent ones and black circles zeros. For instance, the right-most pattern detects corners in an image.

The histograms computed in Eq. (A.1) are rough estimates of the probability density function of the regional data and contain information about the micro-patterns therein. The histogram domain $i$ is a positive integer that corresponds to one of the $n = 2^P$ binary numbers produced by the operator $LBP_{PR}$ in the $P$-pixel neighborhood. The resulting feature vector length of $m \times 2^P$ can be very high, even for typical and modest values of $P$. A typical histogram will group the domain values to reduce variance of the distribution estimate, resulting also in a reduction of the vector size. Ojala *et al.* [23] proposed a clustering procedure based on bit transition counts. They argued that no significant pattern or texture information was contained in bit patterns that had more than two transitions. For example the 8-bit binary strings 11100011 and 00000110 each contain two transitions; string 100000000 contains one transition only. They proposed a code that lumps all the bit patterns with more than two transitions into one bin. Such strings are said to have uniform patterns which comprise nearly 90 percent of all patterns in the (8,1) neighborhood. By accumulating the binary patterns with more than two transitions into a single bin, the $LBP_{PR}^{u2}$ operator can construct a feature vector of dimension much less than $2^P$ bins. The reduction in size for 8 and 16 bit LBP codes is 256 to 59 and for 65536 to 243, respectively.

The feature vector assignment algorithm steps are as follows: a) labeling all the pixels, absent the borders, using the LBP operator, b) dividing the image into $m$ small rectangular regions $R_0, R_1, ..., R_{m-1}$, c) obtaining the histogram of each region, and d) concatenating all the histograms into one vector.

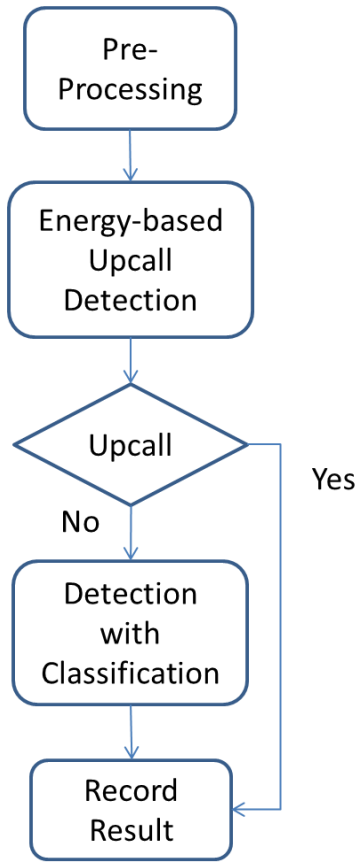

Fig. 1. The proposed call detection scheme

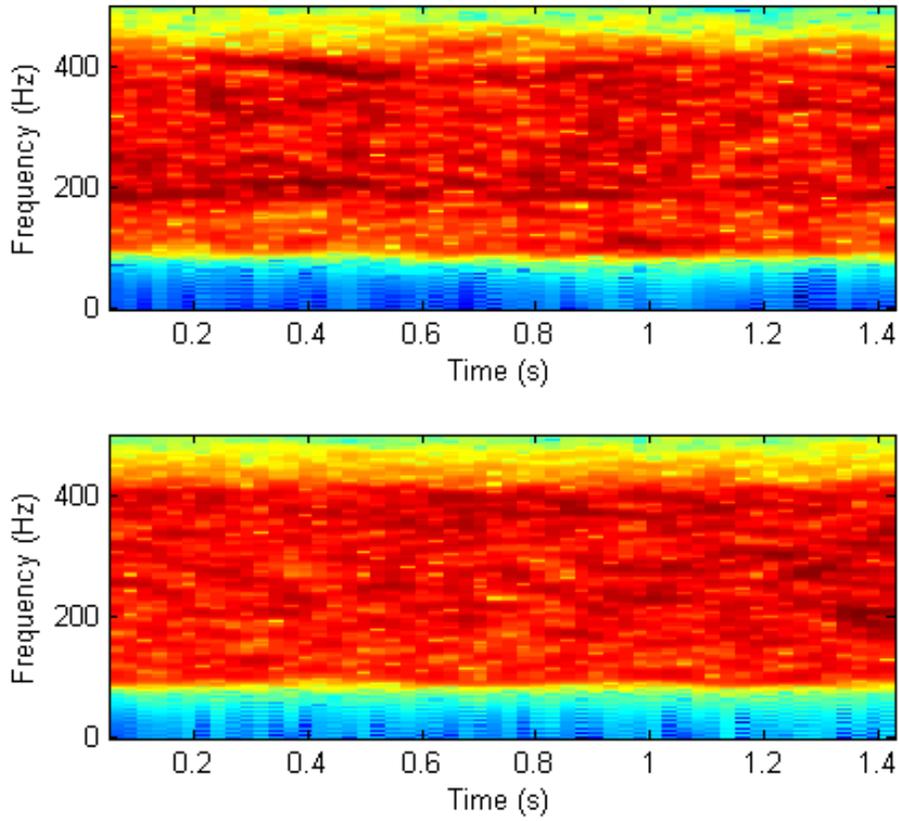

Fig. 2. Two examples of data without any marine mammal calls

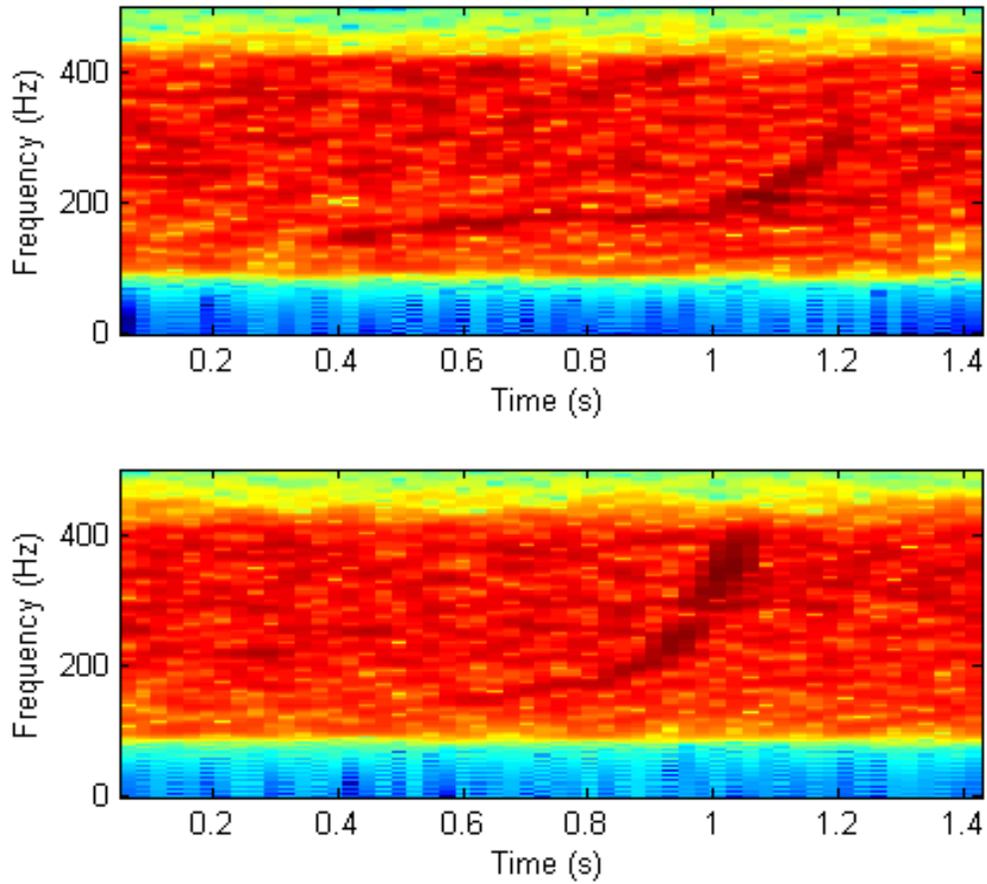

Fig. 3. Two examples of data containing calls very similar to NARW upcalls

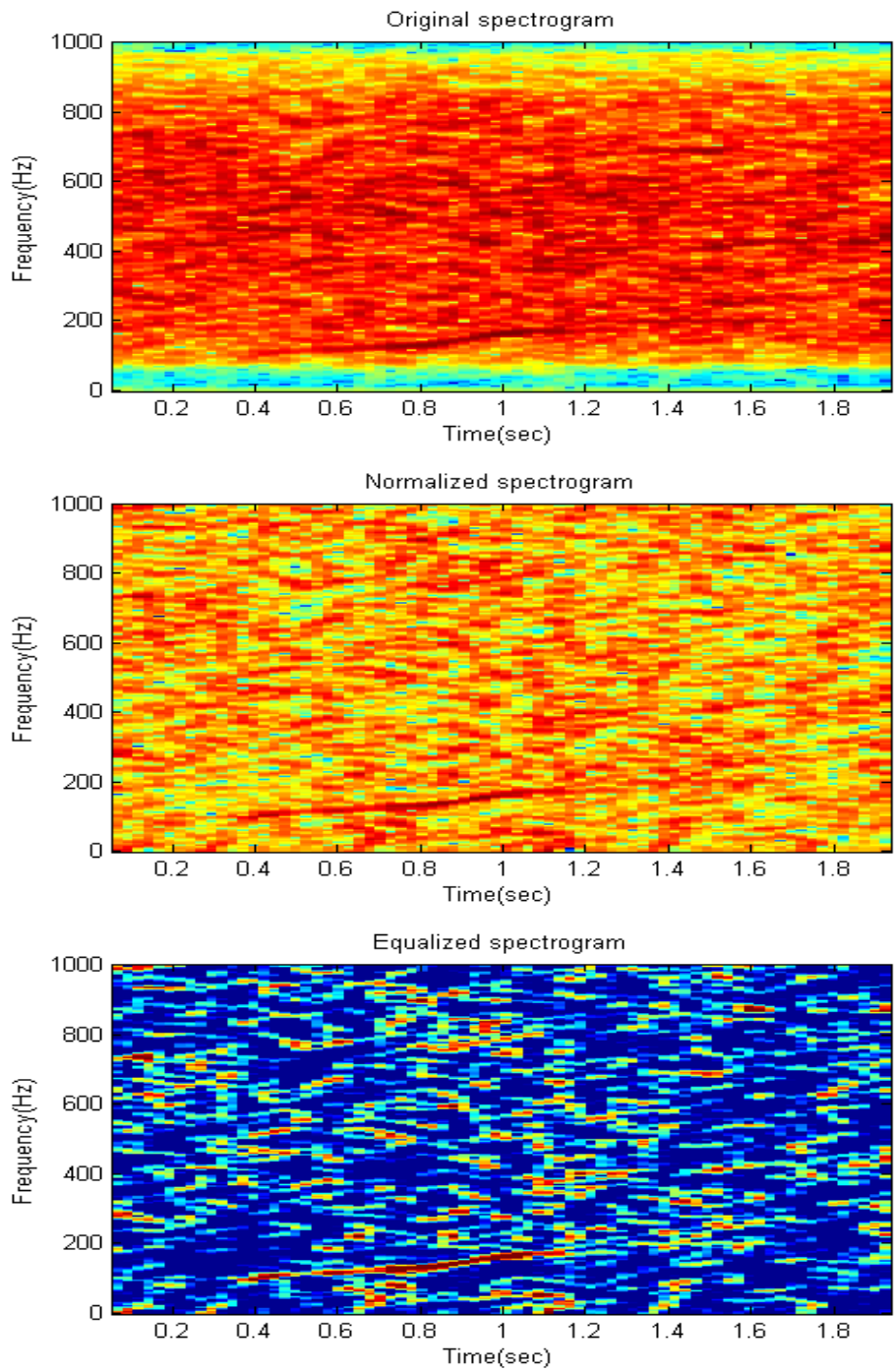

Fig. 4. Original, normalized and equalized spectrograms

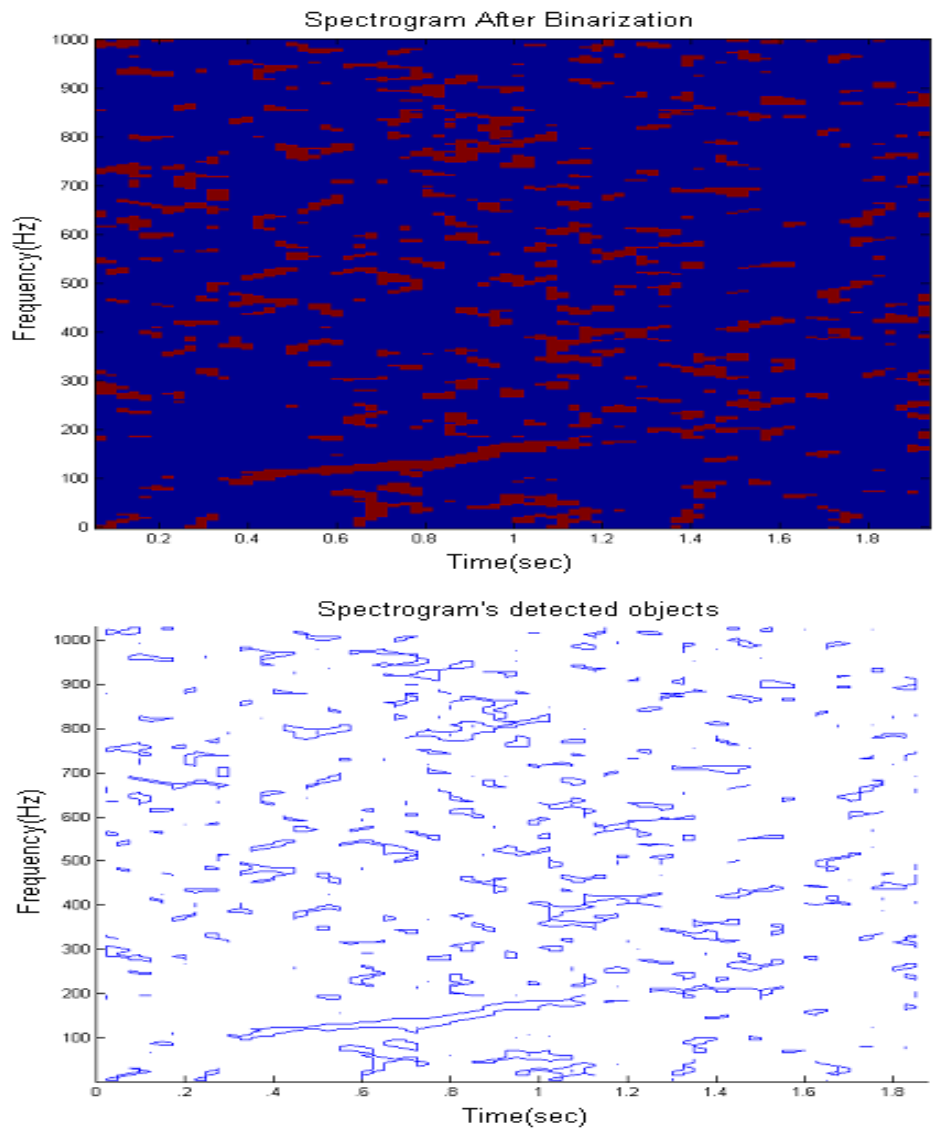

Fig. 5. Spectrogram after binarization and object detection

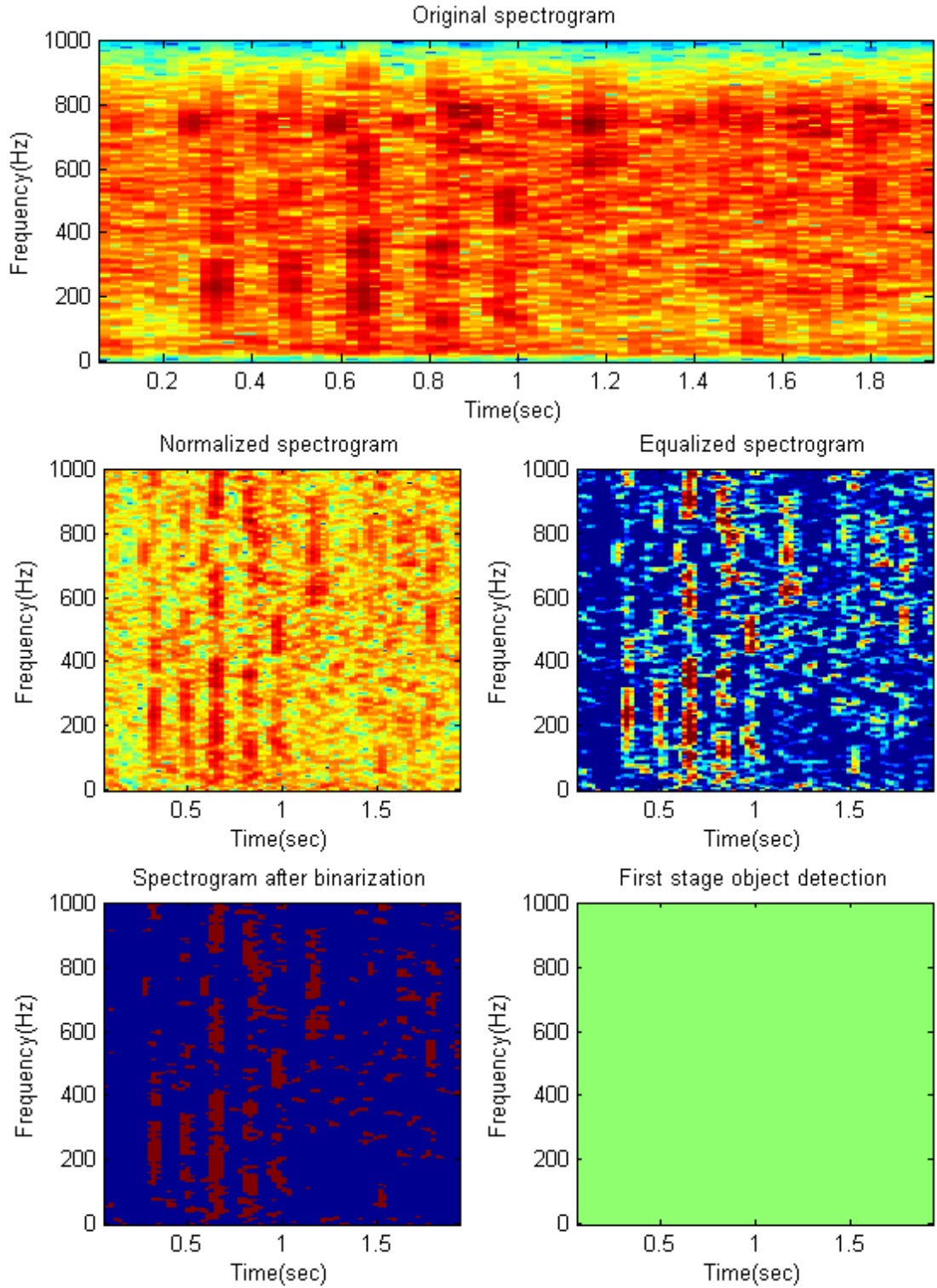

Fig. 6. An example of preprocessing and first stage detection leading to a no-upcall decision

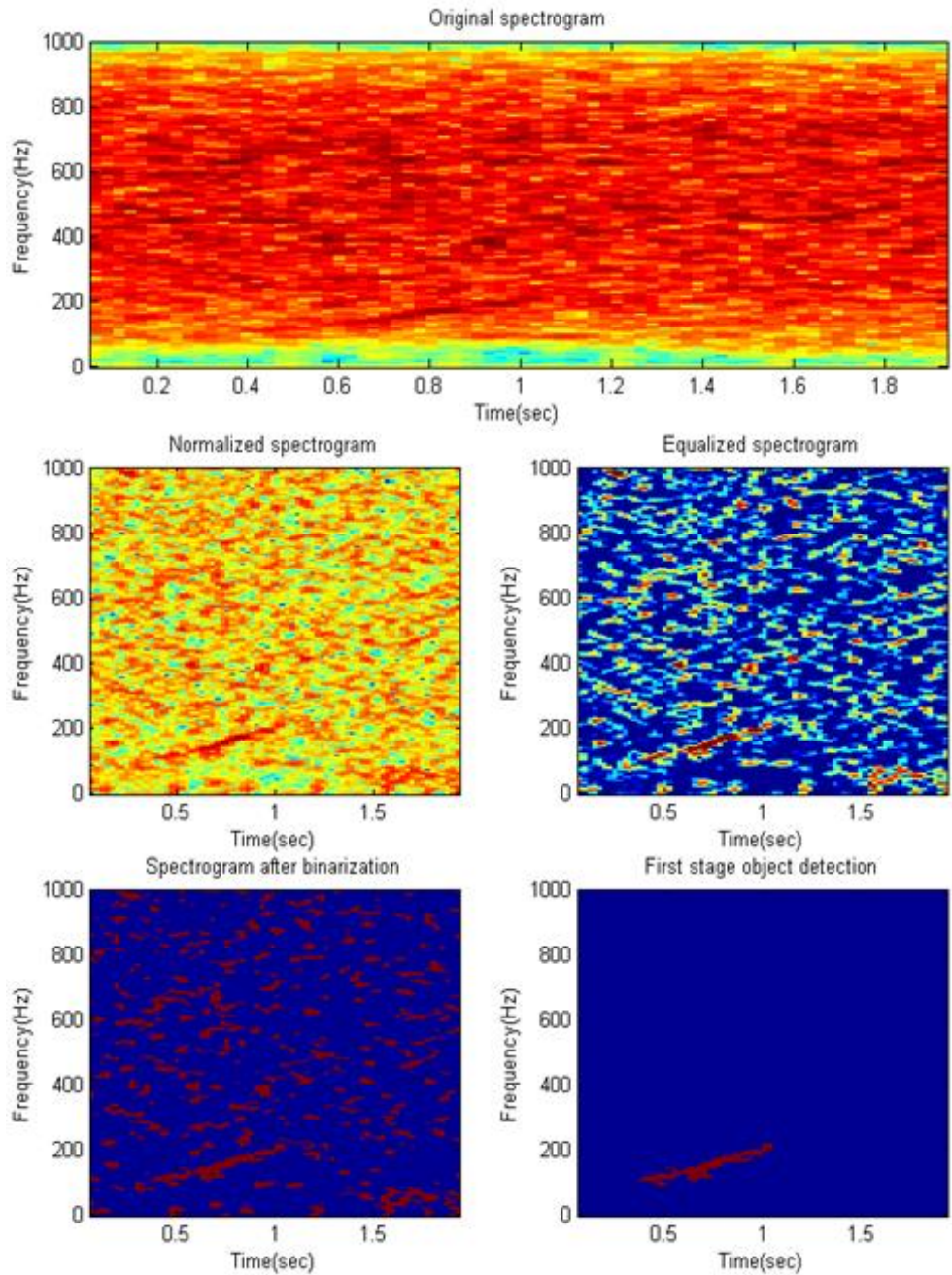

Fig. 7. The outputs of preprocessing steps and first stage detector leading to potential upcall

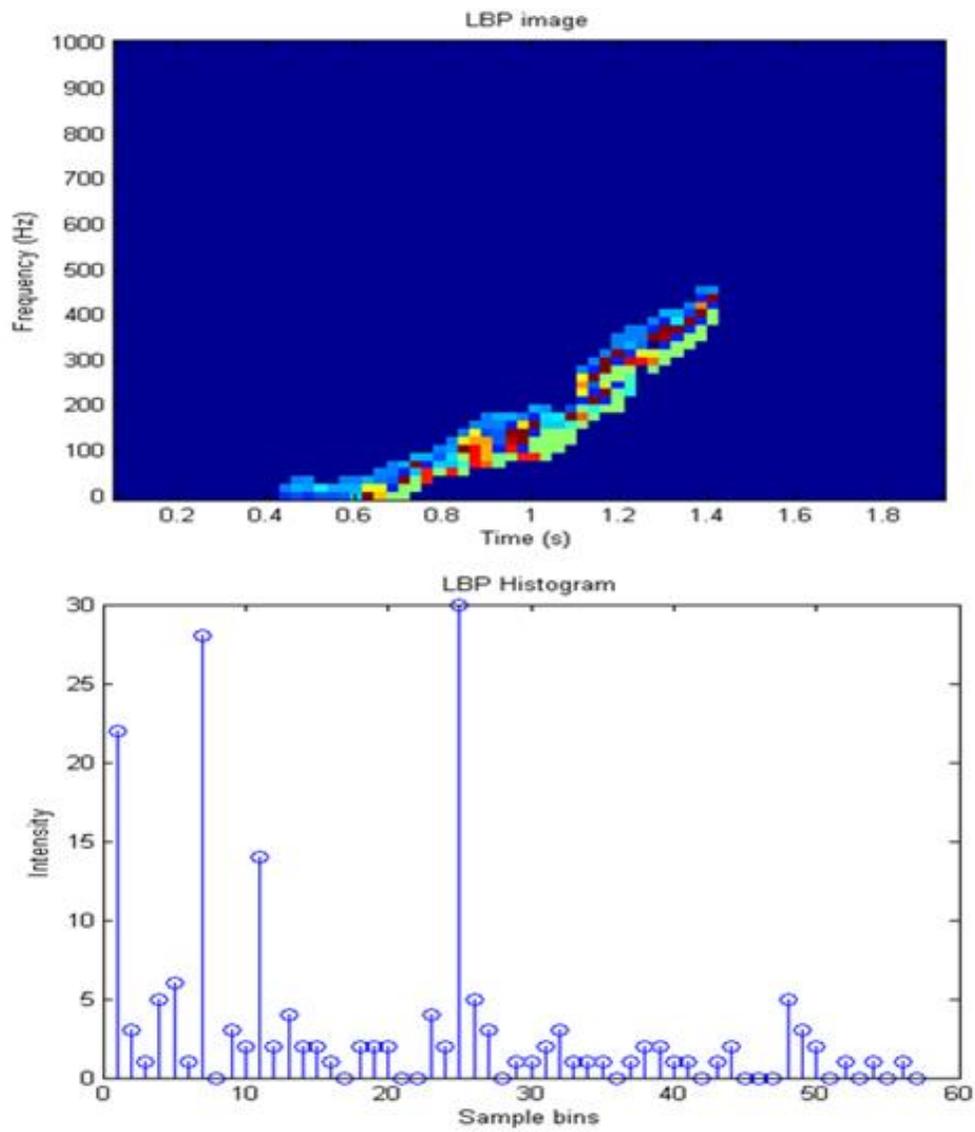

Fig. 8. LBP image (top), LBP histogram (bottom)

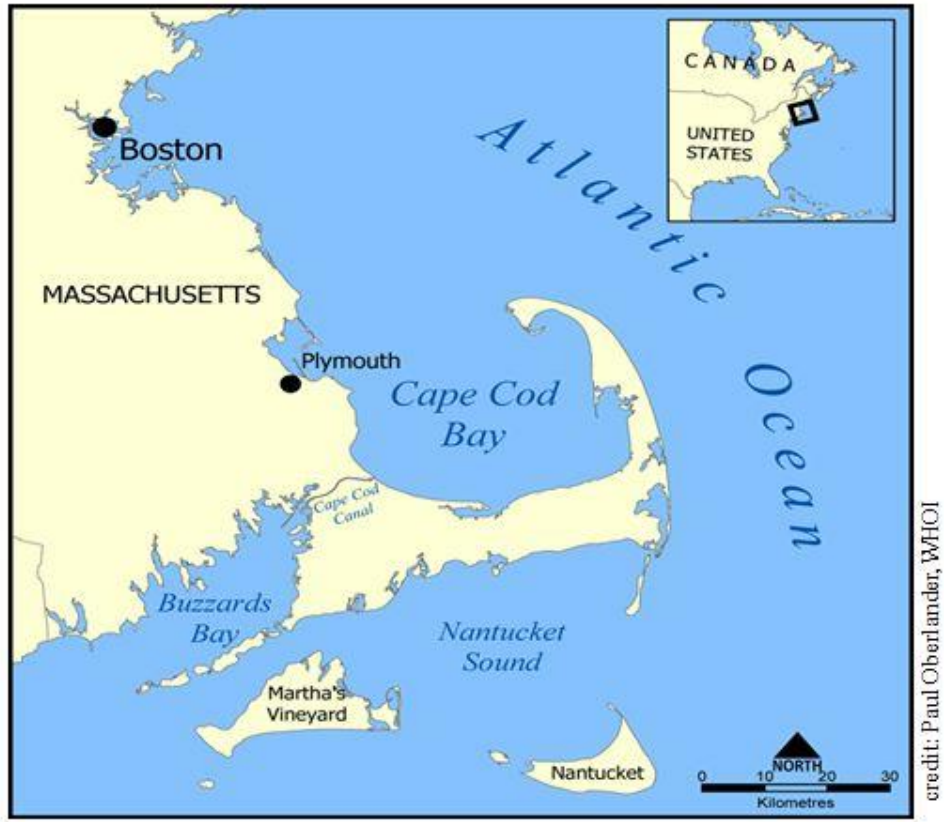

credit: Paul Oberlander, WHOI

Fig. 9. Cape Code Bay map of NARW recordings

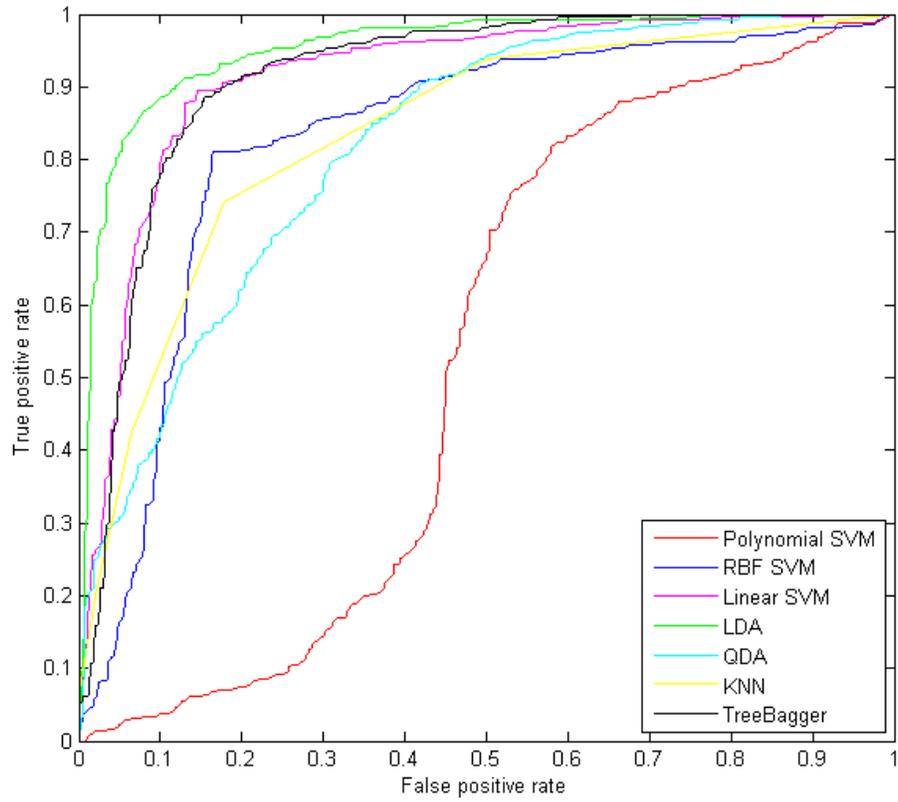

Fig. 10. ROC plot of different classifiers using TFP-2 features

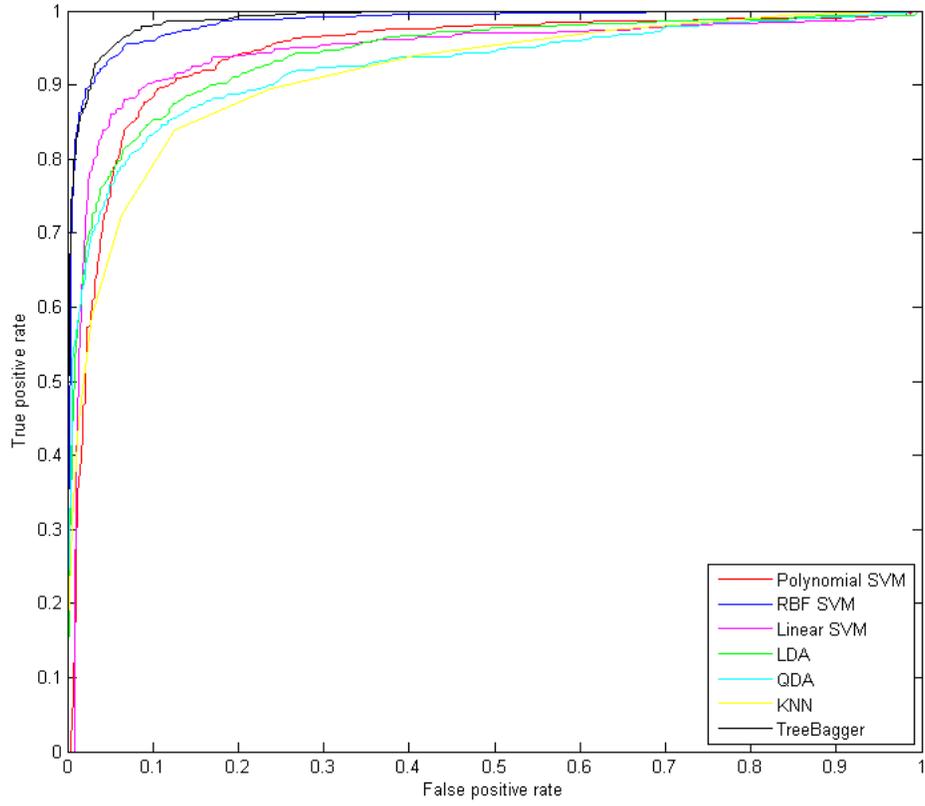

Fig. 11. ROC plot of different classifiers using LBP features

Table 1. Detection results using TFP-2 features

| Classifier | Upcall detection rate (%) | Non-upcall detection rate (%) | Overall detection rate (%) |
|---|---|---|---|
| LDA | 80.11 | 95.21 | 91.7 |
| QDA | 67.38 | 92.78 | 86.87 |
| KNN | 63.23 | 95.74 | 88 |
| Decision Tree | 31.18 | 98.52 | 83.97 |
| Linear SVM | 70.81 | 97.08 | 90.97 |
| RBF SVM | 64.66 | 96.26 | 88.90 |
| Polynomial SVM | 41.05 | 94.04 | 81.70 |
| TreeBagger | 76.25 | 95.83 | 91.27 |

Table 2. Detection results using LBP features

| Classifier | Upcall detection rate (%) | Non-upcall detection rate (%) | Overall detection rate (%) |
|---|---|---|---|
| LDA | 72.96 | 97.82 | 92.03 |
| QDA | 78.40 | 94.44 | 90.70 |
| KNN | 78.97 | 95.35 | 91.53 |
| Decision Tree | 57.79 | 90.65 | 83 |
| Linear SVM | 90.41 | 93.44 | 92.73 |
| RBF SVM | 86.70 | 93.56 | 91.97 |
| Polynomial SVM | 59.23 | 92.57 | 84.80 |
| TreeBagger | 89.98 | 93.48 | 92.67 |

Table 3. AUC values for different classifiers with TFP-2 features

| Classifier | Polynomial SVM | RBF SVM | Linear SVM | LDA | QDA | Tree-Bagger | KNN |
|---|---|---|---|---|---|---|---|
| AUC | 0.618 | 0.8124 | 0.9154 | 0.9528 | 0.82 | 0.9091 | 0.8358 |

Table 4. AUC values for different classifiers with LBP features

| Classifier | Polynomial SVM | RBF SVM | Linear SVM | LDA | QDA | Tree-Bagger | KNN |
|---|---|---|---|---|---|---|---|
| AUC | 0.8838 | 0.9703 | 0.9626 | 0.9534 | 0.9384 | 0.97 | 0.9415 |